\documentclass[aps,twocolumn,prb,10pt]{revtex4-1}
\usepackage{graphicx}
\usepackage{bm}
\usepackage{amssymb}
\usepackage{amsfonts}
\usepackage{amsmath}
\usepackage{amsbsy}
\usepackage{natbib}
\usepackage{comment}
\usepackage[colorlinks=true, urlcolor=blue, citecolor=blue,linkcolor=blue]{hyperref}
\usepackage{color}
\usepackage[T1]{fontenc}
\usepackage{amsthm}
\usepackage{scalerel}

\usepackage{lmodern}
\usepackage{graphicx}
\usepackage{mathtools}

\begin{document}

\title{Interlayer exciton-polaron in atomically thin semiconductors}

% Author: Please give full first and last names for authors and include * after the name of all corresponding authors

\author{M.A. Semina}
\email{msemina@gmail.com}
\affiliation{Ioffe Institute, 194021 St.~Petersburg, Russia}

\author{M.M. Glazov}
\email{glazov@coherent.ioffe.ru}
\affiliation{Ioffe Institute, 194021 St.~Petersburg, Russia}

\author{E.Ya. Sherman}
\email{evgeny.sherman@ehu.eus}
\affiliation{Department of Physical Chemistry, The University of the Basque Country UPV/EHU, 48080 Bilbao, Spain}
\affiliation{IKERBASQUE Basque Foundation for Science, Bilbao, Spain}

\begin{abstract}

A novel type of exciton-phonon bound state -- interlayer polaron -- 
in a double-layer two-dimensional semiconductor with transition metal dichalcogenides as an example, is predicted. 
In these systems the interaction of the 
interlayer exciton with the soft modes of out-of-plane lattice vibrations caused by van der Waals forces  
and flexural rigidity gives rise to a bound quasiparticle. The energy and effective mass
of the formed polaron for weak and strong exciton-phonon coupling regimes are calculated and analyzed. 
Possible manifestations of these effects in transport- and spectroscopy-related experiments are discussed.

\end{abstract}

% Keywords: Please provide a minimum of three and a maximum of seven keywords, separated by commas


\maketitle

\section{Introduction}

Coupling between lattice vibrations and electrons in solids produces a great variety of experimentally observable 
effects, usually described as ``the physics of polarons''. Different regimes of this coupling build
weak and strong-coupling polarons influencing carriers mobility, effective masses, and response functions of the system \cite{Devreese2007}.
{Polaron} effects are critically important for excitons \cite{RSE} produced by light absorption since
{composite} structure of the excitons enhances the variety of the {phenomena}
related to their coupling to the host lattice (e.g., Ref. \onlinecite{Kusmartsev1983}). 
In optical properties, the exciton-phonon coupling can lead to
the temperature-dependent light absorption threshold \cite{Ioselevich1982}, known as the Urbach rule.

Recent progress in experimental and  theoretical studies of the transition metal dichalcogenide (TMDC) mono-  
and bilayers demonstrated that a coupling between the carriers and the lattice can have a strong impact on the properties 
of these materials. Electron- and exciton-phonon interaction results in pronounced sidebands 
in absorption and emission spectra~\cite{PhysRevLett.119.187402,shree2018exciton}, controls resonant 
light scattering and impacts coherence generation~\cite{PhysRevLett.115.117401,PhysRevResearch.1.032007,He:2020aa,PSSR:PSSR201510291}. 
The electron-phonon interaction serves as an efficient channel of the momentum and energy relaxation of charge 
carriers and excitons~\cite{PhysRevB.85.115317,Song:2013uq,PhysRevB.90.045422}. In general, polaron effects related 
to the coupling of the charge carriers with in-plane polarized phonons are important for 
the physics of TMDC monolayers~\cite{doi:10.1063/1.5025907,Xiao_2017,PhysRevB.98.045143,PhysRevB.100.041301}. 

{A remarkable property of two-dimensional crystals is the presence of soft flexural phonon modes 
\cite{ll7_eng,nelson:2004aa,Falkovskii:2012aa,Katsnelson:2013aa}. These out-of-plane vibrations are 
responsible for rippling and crumpling of the two-dimensional materials as well as for their 
anomalous elastic properties~\cite{PhysRevB.92.155428,Gornyi:2016aa,2019arXiv190702010A}. 
Flexural modes demonstrate low damping
with significant potential for mechanical and 
optomechanical applications~\cite{Morell:2016aa,Morell:2019aa}.  However, as compared to the in-plane modes, 
the interaction of charge carriers with the flexural vibrations is diminished and, at first glance, 
the polaron effects are weak. Here we show that this is not the case for excitons in bilayer TMDCs, 
where the substantial polaron coupling with the out-of-plane modes is expected. Indeed, the modulation 
of the interlayer distance by the flexural phonons results in variations of the Coulomb energy of 
electron-hole interaction and provides a significant impact on the interlayer exciton.} 
 The special 
properties of this interlayer exciton-polaron are related 
to the fact the Coulomb attraction between electron and hole tends to decrease the interlayer 
distance and the softness of the flexural modes makes such coupling particularly efficient. 
Thus,  a large-size polaron with the spatial dimension exceeding the exciton Bohr 
radius, typically of the order of few nanometers, can be formed.

{The excitonic effects in bilayer TMDCs attract nowadays an increasing interest due to manifestations of the novel 
physics related to the combinations of layer and valley degrees 
of freedom~\cite{Song:2018aa,Mak:2018aa,Jin:2018aa,Rivera:2018aa,Seyler:2019aa,Tran:2019aa,Jin:2019aa,Alexeev:2019aa}. 
The analysis presented in this paper shows that the formation of the interlayer exciton-polarons is an essential element of the
bilayer TMDC physics and, therefore, should be taken into account while analyzing the properties of excitons in such structures. }

In this paper we theoretically study the polarons formed by excitons causing lattice deformation in 
TMDC two-layer structures in the realizations from weak to strong coupling regimes. 
In addition to the composite structure of the exciton, as related to its Bohr radius, these 
regimes are determined by the interplay of the interlayer coupling due to the van der Waals forces and the flexural rigidity of 
the monolayers. We calculate the energy and effective mass of the polarons in both these regimes 
and analyze possible implications of the interlayer exciton-polaron formation relevant for experimental observations.

\section{Model: bilayer exciton-phonon coupling}

\subsection{Flexural phonon modes}

\begin{figure}[h]
\includegraphics*[width=0.95\columnwidth]{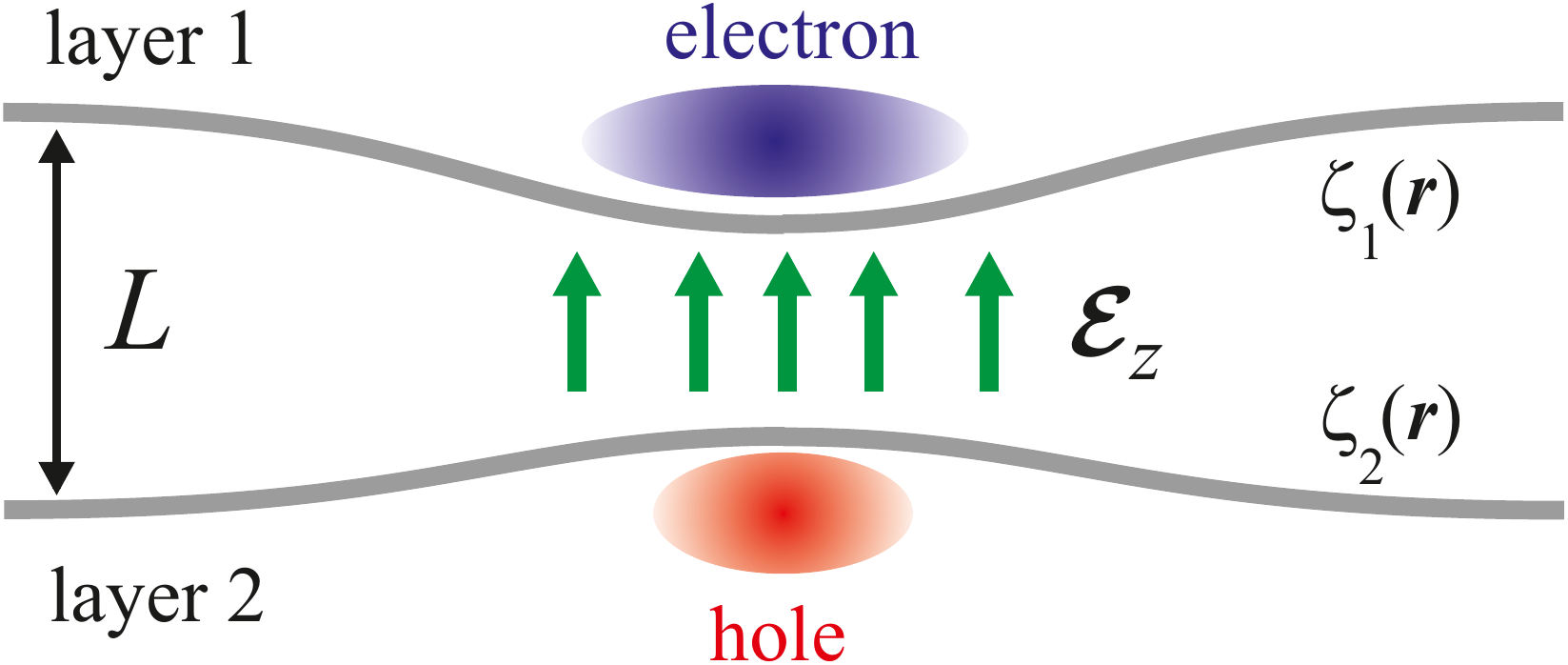}
\caption{Schematic illustration of the bilayer exciton-polaron. {Here the space between the monolayers is not filled by any material and} 
attraction between electron and hole described 
by the electric field ${\cal E}_{z}$ leads to deformation of the monolayers and formation of a nanosize polaron
state.} 
\label{fig:scheme}
\end{figure}

The studied system {hosting an interlayer exciton,} consists of two monolayers (ML{s}), separated by the distance $L,$ as depicted in {Figure \ref{fig:scheme}}. 
For simplicity we assume that the monolayers are identical, but electrically biased so that the top layer 
is preferably occupied by electrons and the bottom one by the holes. Thus, the interlayer exciton represents a dipole 
polarized along the structure normal ${\bm z}.$ The variation of the interlayer distance due to flexural vibrations modulates the electron 
hole separation and provides efficient exciton-phonon coupling. {We assume that there 
is no material sandwiched between the layers, similarly to the structures studied in Refs. \onlinecite{Morell:2016aa,Morell:2019aa}.}

{First, we consider the system of two freestanding electrically neutral monolayers,} Figure ~\ref{fig:scheme}, characterized by 
the out-of-plane displacements $\zeta_{1}({\bm r})$ and $\zeta_{2}({\bm r}).$ 
{Their motion is} described by the coupled equations with indices $i,j=1,2$ numerating the monolayers as [cf. Ref. \onlinecite{ll7_eng}]:
\begin{equation}
\label{2ML:forces}
\varrho \frac{\partial^2 \zeta_{i}({\bm r})}{\partial t^2} + 
B\Delta^2\zeta_{i}({\bm r})
= 
\varrho\frac{\omega_0^2}{2} (\zeta_{j}({\bm r}) - \zeta_{i}({\bm r})), 
\end{equation}
where $t$ is the time, $\Delta$ is the Laplace operator acting in the ${\bm r}$-space, 
$\varrho$ is the two-dimensional mass density (mass per unit area per monolayer), and $B$ 
is the flexural rigidity of the monolayer. Although in this paper we consider for simplicity 
identical layers, similar results hold also for the layers of different materials {and for the case {where} one of the layers is attached to 
a substrate.}
The interlayer coupling parameter $\omega_{0}$ {can be} related to, e.g., the van der Waals forces. At negligible ($\omega_{0}=0$) 
coupling {substitution of a plane wave 
$\zeta_i (\bm r) = \zeta_{\bm q, \omega_q}^{(i)} \exp{(\mathrm i \bm q \bm r - \mathrm i \omega_q t)}$ into Equation  ~\eqref{2ML:forces} 
results in two independent algebraic equations  for the Fourier-components of displacements, 
$\zeta_{\bm q, \omega}^{(i)}$, in the form $\left(-\omega_q^{2}\rho+ B q^4\right)\zeta_{\bm q, \omega_q}^{(i)}=0$, where $\bm q$ 
is the in-plane phonon wavevector. As this equation has nonzero solutions only if $Bq^{4}=\omega_q^{2}\rho,$}
we obtain degenerate flexural modes with the dispersion   
\begin{equation}
\label{1ML:omega}
\omega_{q} = \varkappa q^{2}, 
\end{equation} 
where  we introduced $\varkappa^2\equiv B/\varrho.$ 
Note that at small $q,$ the dispersion in \eqref{1ML:omega} 
is much softer than that of the {in-plane polarized} acoustic phonons, linear in the wavevector $q.$ 
{The presence of soft $\omega_q \propto q^2$ modes is a specific
feature of two-dimensional systems and membranes~\cite{ll7_eng,nelson:2004aa,Falkovskii:2012aa,Katsnelson:2013aa},
resulting from the $\propto\Delta^2\zeta_{i}({\bm r})-$term in the dynamical equation.} 

{In the general case of $\omega_0\ne 0$, it follows from \eqref{2ML:forces}, that there are two different in the symmetry and in the {dispersion} 
modes of the out-of-plane vibrations, which can be characterized as the in-phase vibrations with $\zeta_{1}({\bm r}) = \zeta_{2}({\bm r})$
and the out-of-phase breathing mode with $\zeta_{1}({\bm r}) = -\zeta_{2}({\bm r}).$ Since the in-phase mode is odd with respect to the reflection 
in the symmetry plane of the bilayer and conserves the distance between electron and hole, it cannot significantly affect the 
interlayer exciton in the harmonic approximation {in contrast to the breathing mode. The out-of-phase vibrations are described} by the modified equation 
\begin{equation}
\label{1ML:-}
\frac{\varrho}{2} \frac{\partial^2 \zeta({\bm r})}{\partial t^2} + 
\frac{\varrho}{2} \varkappa^2 \Delta^2\zeta({\bm r}) +
\frac{\varrho}{2}\omega_0^2\zeta({\bm r}) = f(\bm r),
\end{equation}
where $\zeta({\bm r}) = \zeta_{1}({\bm r}) - \zeta_{2}({\bm r})$ is the variation of the interlayer distance,  and
$f(\bm r)$ is the external local pressure applied here at the layer 1 (layer 2) due the electron-hole attraction corresponding to the product of the 
contribution of hole (electron) to the electric field ${\cal E}_{z}$ 
in Figure  \ref{fig:scheme} and the local electron (hole) charge density (see Appendix \ref{app:form-factor} for more details).  
To specify the explicit form of $f(\bm r)$ one needs the mechanism of exciton coupling 
to the out-of-phase breathing mode, which will be defined in next subsection.}

{Note that in absence of the exciton, $f(\bm r)=0$, 
setting  $\zeta(\bm r) = \zeta_{\bm q, \omega_q}\exp{(\mathrm i \bm q \bm r - \mathrm i \omega_q t)}$ 
we obtain the algebraic equation $(-\omega_q^2 + \varkappa^2 q^4 + \omega_0^2)\zeta_{\bm q, \omega_q}=0$.}
The resulting dispersion takes the form
\begin{equation}
\label{2ML:omega}
\omega_q = \sqrt{\omega_0^2 + \varkappa^2 q^4} \approx
\begin{cases}
\omega_0 + {\varkappa^2 q^4}/({2\omega_0}), \quad \varkappa q^2 \ll \omega_0,\\
\\
\varkappa q^2, \quad \varkappa q^2 \gg \omega_0.
\end{cases}
\end{equation} 
At small wavevectors $\omega_{q}$ tends to a constant 
with a quartic in $q$ correction, while at the sufficiently large wavevectors we recover parabolic 
dispersion of the flexural modes in  \eqref{1ML:omega}, {still valid for any $q$ for the in-phase mode here.}

\subsection{Exciton-phonon coupling: what is special for bilayers?}

Now we derive the exciton-phonon interaction in the bilayer structure. 
We take the exciton wavefunction in the form
\begin{equation}
\label{exc:wave}
\Psi_{\bm K} = \frac{\exp{(\mathrm i \bm K \bm R)}}{\sqrt{\mathcal S}} \varphi(\bm \rho),
\end{equation}
where $\mathcal S$ is the normalization area, $\bm K$ is the translational 
wavevector of the exciton, and $\bm R$ and $\bm \rho$ are the in-plane center 
of mass and relative motion coordinates defined as:
\begin{equation}\label{Rrho}
\bm R = \frac{m_{e}}{m} \bm \rho_{e} + \frac{m_{h}}{m} \bm \rho_{h}, \qquad \bm \rho = \bm \rho_{e} - \bm \rho_{h},
\end{equation}
with $m=m_{e}+m_{h}$ being the exciton translational mass. {Hereafter we assume that the bilayer 
system is translationally invariant and neglect moir\'e effects related to the possible incommensurability 
of the lattices in the top and bottom layers as well as the twist angle~\cite{Seyler:2019aa,Tran:2019aa,Jin:2019aa,Alexeev:2019aa}. 
These effects are most prominent for the layers in a tight contact, 
which enables efficient hybridization of the electron orbitals. The moir\'e effects produce an additional potential acting on the 
excitons~\cite{PhysRevLett.118.147401,PhysRevB.97.035306} and affects their optical properties. The period 
of the potential is controlled by a twist angle and is typically sufficiently large as compared to the Bohr exciton 
and polaron radii calculated below, making it possible to consider system as a translationally 
invariant and effectively homogeneous on the length scales related to the polaron, particularly, if the layers 
are sufficiently far apart, which is of our interest in this paper. 
Thus, at the first approximation, the moir\'e effects can be disregarded.}

The smooth envelope of the relative electron-hole motion $\varphi(\bm \rho)$  can be calculated after 
Refs. \onlinecite{PhysRevB.97.195452,2018arXiv180106310G,PhysRevB.98.115104,Kezerashvili:2019aa,Semina:2019aa}. 
{It satisfies the Schr\"odinger equation
\begin{equation}
\label{schr:rel}
-\frac{\hbar^2}{2\mu} \Delta \varphi(\bm \rho) + V(\bm \rho,z)\varphi(\bm \rho)= E \varphi(\bm \rho),
\end{equation}
with $\mu  = m_{e}m_{h}/m$ being the electron-hole pair reduced mass and $E$ being the eigenenergy. In Equation~\eqref{schr:rel} we have introduced the 
}
the electron-hole attraction potential energy $V(\bm \rho,z)$ 
with $z$ being the separation of the charges along the sample 
normal. The latter is typically given by the 
Coulomb or extensions of Rytova-Keldysh potential~\cite{Semina:2019aa,Rytova1967,1979JETPL..29..658K}. 
At a fixed interlayer distance $z=L$, Equation~\eqref{schr:rel} gives the series of the interlayer exciton states. 
It also permits to calculate the exciton-phonon interaction taking into account a variation of the exciton potential 
energy due to the variation $\delta z$ of the interlayer distance $z$. The latter occurs as a result of the phonons. 
Taking into account that the variation of the potential energy in the linear-in-displacement order is given by
\begin{equation}
\delta V = \left. \frac{\partial V(\bm \rho,z)}{\partial z}\right|_{z=L} \delta z 
= \left. \frac{\partial V(\bm \rho,z)}{\partial z}\right|_{z=L} [\zeta_{1}(\bm \rho_{e}) - \zeta_{2}(\bm \rho_{h})],
\end{equation}
we arrive at the following 
expression for the exciton-phonon interaction matrix element
\begin{eqnarray}
&&\hspace{-0.5cm}U_{\bm K'\bm K} = \int d^{2}R d^{2}\rho\, \Psi_{\bm K'}^{*}\Psi_{\bm K} V_{z}(\rho,{L})
[\zeta_{1}(\bm \rho_{e}) - \zeta_{2}(\bm \rho_{h})] \\
&&=\frac{1}{2}\int d^{2}R d^{2}\rho\, \Psi_{\bm K'}^{*}\Psi_{\bm K} 
V_{z}(\rho,{L})[\zeta(\bm \rho_{e}) + \zeta(\bm \rho_{h})], \nonumber
\label{U:rspace:1}
\end{eqnarray}
where, according to \eqref{Rrho}, $\bm \rho_{e} = \bm R +(m_{h}/m) {\bm\rho}$ and $\bm \rho_{h} = \bm R - (m_{e}/m){\bm\rho},$
we have taken into account that $\zeta_{1}(\bm \rho_{e})=\zeta(\bm \rho_{e})/2$ and $\zeta_{2}(\bm \rho_{h})=-\zeta(\bm \rho_{h})/2,$
and introduced notation for the $z$-{component of the} effective field as the derivative $V_{z}(\rho,z)\equiv{\partial V(\rho,z)}/{\partial z}$.

Quantizing the phonon-induced displacements in a standard way {(note that $\zeta({\bm\rho})$ is the difference of the layers displacements)}
\begin{equation}
\label{displ:qnt}
\zeta(\bm \rho) = \sum_{\bm q} \sqrt{\frac{\hbar}{\varrho \omega_q \mathcal S}} 
b_{\bm q}^\dag e^{-\mathrm i \bm q \bm \rho} +{\rm h.c.},
\end{equation}
with $b_{\bm q}^\dag$ ($b_{\bm q}$) being the phonon creation and 
annihilation operators {and transferred wavevector} {$\bm q = \pm (\bm K' - \bm K)$ depending whether the phonon absorption/emission is studied},  
we obtain the following exciton-phonon coupling Hamiltonian
\begin{equation}
\label{int}
U = \sum_{\bm q} U_{\bm q} e^{\mathrm i \bm q {\bm R}} b_{\bm q}c_{\bm K+\bm q}^\dag c_{\bm K} + {\rm h.c.}, 
\quad U_{\bm q} = D{\mathcal F}_{s}(q)\sqrt{\frac{\hbar}{\varrho \omega_q \mathcal S}},
\end{equation}
{where $c_{\bm K+\bm q}^\dag$ and $c_{\bm K}$ are the exciton creation and annihilation operators, respectively.} 
Here the deformation potential constant of {the} exciton-phonon coupling
\begin{equation}
\label{classical:int}
D = \begin{cases}
{4e^{2}}/({\varepsilon a_{B}^{2}}), \quad L \ll a_B,\\
\\
{e^{2}}/({\varepsilon L^{2}}), \quad L \gg a_B,
\end{cases}
\end{equation}
where $\varepsilon$ is the effective dielectric constant, $a_{B}$ is the interlayer exciton Bohr radius, 
and ${\mathcal F}_{s}(q)$ is the corresponding form-factor with ${\mathcal F}_{s}(0)=1,$
as specified in Appendix \ref{app:form-factor}.
 
{Before proceeding to the next section, where we use Equation ~\eqref{int} to calculate the polaron states, we mention that the proposed mechanism 
of exciton-phonon coupling is qualitatively different from that for single-layer excitons. This coupling is directly based on phonon-induced 
modulation of the Coulomb interaction binding the exciton. However, the intra-layer exciton-polaron coupling is very weak 
due to the inefficiency of the Fr\"{o}hlich interaction in two dimensions. Another reason for this weakness is the similarity of
electron and hole effective masses producing a low net charge density inside the exciton and, thus, making it weakly sensitive to the phonon-induced fields 
(e.g., Ref. \onlinecite{Kusmartsev1983}). 
The coupling to the in-plane acoustic phonons via deformational potential is also quite weak, mainly, because the matrix element vanishes at small 
momenta.}

\section{Results}

\subsection{Weak coupling}

\label{sub:weak}

Let us begin with the standard perturbative analysis of the interlayer polaron. 
{In the weak coupling regime, the polaron shift of the exciton 
ground state evaluated at a temperature $T$ in the second-order perturbation theory is presented as
\begin{eqnarray}
\label{pert:1:T}
&&\delta E_{w} = - \sum_{\bm q} |U_q|^2\times \\
&&\left[\frac{1+n_q}{\hbar\omega_q + \hbar^{2}q^{2}/(2m)} + \frac{n_q}{\hbar\omega_q - \hbar^{2}q^{2}/(2m)}\right], \nonumber
\end{eqnarray}
where $n_q = [\exp(\hbar\omega_q/T)-1]^{-1}$ is the phonon distribution function, and the terms $\propto(1+n_q)$ and $\propto\,n_{q}$ in the square 
brackets describe emission and absorption of the phonons, respectively. 
To elucidate the main contributions to the exciton-phonon coupling constant, we begin with the $T=0$ case. }
Since for the realizations of interest, ${\mathcal F}_{s}(q)$ 
is close to one, as shown in Appendix \ref{app:form-factor} {and also discussed below},
we {omit} it in Equation  \eqref{pert:1:T}. Therefore, the sum over $\bm q$ can be readily evaluated as
\begin{eqnarray}
\label{pert:2}
\delta E_{w}&=& - \frac{D^2}{\varrho} \frac{\tanh^{-1}(\sqrt{1-\varkappa^2/\mathcal K^2})}{4\pi \mathcal K\sqrt{{1}-\varkappa^2/\mathcal K^2}\omega_0} \\
&&\approx -\frac{\beta\hbar\omega_0}{2\pi}\ln{\sqrt{\frac{\mathcal K}{\varkappa}}}.\nonumber
\end{eqnarray}
Here {the effective coupling constant $\beta$ reads:}
\begin{equation}
\label{beta}
\beta = \frac{2m D^{2}}{\varrho\hbar^{2}\omega_{0}^{2}},
\end{equation}
{the parameter $\mathcal K= \hbar/(2m)$ characterizes steepness of the exciton dispersion}, and the last approximate equality holds for $\mathcal K \gg \varkappa$. 
{At a relatively high temperature $T\ge\hbar\omega_{0}$ we obtain from Equations \eqref{pert:1:T} and \eqref{beta} 
that the constant $\beta$ is enhanced by a factor $2T/(\hbar\omega_{0}).$}
Since the flexural elasticity of the layers is determined by the bending of chemical bonds \cite{Casillas2014}, it can be estimated
as $\varrho\varkappa^{2}\sim \hbar^{2}/(m_{{0}}a_{0}^{2}),$ where $a_{0}$ is the in-{plane} lattice constant {and $m_0$ is the free-electron mass.} 
Taking into account that {the exciton effective mass $m\sim m_0$ in TMDC-based systems~\cite{RevModPhys.90.021001} and} 
estimating $\varrho\sim M/a_{0}^{2}$ through the unit cell mass $M$, we obtain $\varkappa\sim \hbar/\sqrt{Mm}$. 
Therefore, ${\mathcal K}/\varkappa\sim \sqrt{M/m}\gg 1,$ being determined by the inverse of the adiabatic parameter in the theory of electron-phonon coupling.
{Next, we can estimate the van der Waals related frequency in the following way. 
First, we take into account that the van der Waals interaction 
energy of two units cells (separated by the in-plane distance $d$) in the TMDC layers, $w_{W},$ 
can be estimated as \cite{lp9_eng} $w_{W}\sim -\hbar^{2}/(m_{0}a_{0}^{2})\times a_{0}^{6}/\left(L^{2}+d^{2}\right)^{3}.$
Summing over the units cells in a single layer, we obtain the total interaction energy of the given 
unit cell with the other layer as $W_{W}\sim -\hbar^{2}/(m_{0}a_{0}^{2})\times a_{0}^{4}/L^{4}.$ As a result, we estimate the energy $\hbar\omega_{0}$ as
$\sim\hbar\sqrt{|d^{2}W_{W}/dL^{2}|/M}\sim\hbar^{2}/(m_{0}a_{0}^{2})\times\sqrt{m/M}\times a_{0}^{3}/L^{3}.$ } 

In the same $\mathcal K\gg \varkappa$ limit the effective polaron mass of the exciton {at $T=0$} reads~\cite{PhysRevB.100.041301}
\begin{equation}
\label{pert:M}
m^*=m\left(1+\frac{\beta}{4\pi} \right).
\end{equation}
The logarithmic enhancement of the polaron shift is a feature of two-dimensional systems [cf. Ref. \onlinecite{PhysRevB.100.041301}]. 
Noteworthy, the sum over $\bm q$ in \eqref{pert:1:T} diverges in the absence of flexural rigidity where $\varkappa \to 0$. 
The presence of the flexural rigidity cuts the divergent sum off at $Q \sim \sqrt{\omega_0/\varkappa}$ 
estimated as $Q\sim a_{B}^{-1}\times\left(\hbar\omega_{0}/E_{B}\right)^{1/2}\times\left(m/M\right)^{1/4}.$
{Thus, depending on the particular realization of the system, the cut-off can be due 
to the flexural rigidity or due to the formfactor ${\mathcal F}_{s}(q).$ 
The inclusion of the form-factor calculated in Appendix \ref{app:form-factor} results only in the change of the 
argument of large logarithm in Equation ~\eqref{pert:2}.}

Importantly, in the absence of the interlayer coupling ($\omega_{0} =0$) the polaron shift 
and mass diverge also. This divergence is caused by the softness of flexural modes: even infinitely 
small external force results in macroscopic deformation of the membrane~\cite{ll7_eng} since the energy
of the flexural deformation of the lateral size ${R}$ behaves as $\varrho\varkappa^{2}\zeta^{2}/{R}^{2}$
and vanishes in the large ${R}-$limit, permitting any finite $\zeta$ for zero energy cost.

\subsection{Strong coupling: polaron formation}

The van der Waals interaction rapidly falls with increasing the distance $L$ between the monolayers \cite{lp9_eng}
leading to $\omega_{0}^{2}\propto L^{-6}$ dependence,
while the deformation potential $D$ decreases rather smoothly, as  
$\propto L^{-2}$ (at $L \gg a_B$). As a result, the coupling parameter $\beta$ can 
be made sufficiently large rendering perturbation theory inapplicable.

Thus, we explore the interlayer polaron in the strong coupling approximation.
{ In this regime, the exciton produces significant deformations of the layers such 
that the effective potential well is formed, which causes the exciton localization. 
Thus, the polaron energy can be recast in the form \cite{Devreese2007,landau:pekar}}
\begin{equation}
\label{var:1}
\delta E_s = \left\langle \psi(\bm r) \left| \frac{\hbar^2 k^2}{2m} \right| \psi(\bm r)\right\rangle - \sum_{\bm q} \frac{|U_q|^2}{\hbar\omega_q} F^2(\bm q).
\end{equation}
Here $\psi(\bm r)$ is the exciton {wavefunction}  in the effective potential well formed by layer displacements, 
to be found by minimization of Equation ~\eqref{var:1}, and
\[
F(\bm q) = \left\langle \psi(\bm r) \left| e^{\mathrm i \bm q\bm r} \right| \psi(\bm r)\right\rangle .
\]
Equation~\eqref{var:1} has a clear physical meaning: The first term is the kinetic energy 
of the exciton in the potential well, and the second term is the gain in the potential energy due to the 
relative displacements of the layers and corresponding decrease in the negative Coulomb energy.  
The latter can be written in the following equivalent form
\begin{equation}
\label{potential}
{E_{\rm pot}} = -\frac{1}{2} \sum_{\bm q} f_{\bm q} \zeta_{\bm q},
\end{equation}
where  $f_{\bm q} = D F(\bm q)/\sqrt{\mathcal S}$ is the Fourier component of the force density due to the presence of exciton, 
while the relation between the force 
and displacement takes the standard form for a harmonic 
oscillator $\zeta_{\bm q} = 2 f_{\bm q}/(\varrho\omega_q^2)$ as it follows from the Fourier transform of Equation ~\eqref{1ML:-}. 
{As we see from the following, if the coupling parameter $\beta$ is small, there is no bound state described 
by Equation ~\eqref{var:1} and the strong-coupling polaron is absent.}

We use {variational approach to minimize the energy in Equation  \eqref{var:1} and choose} the Gaussian form of the trial function 
\begin{equation}
\label{Gauss}
\psi( r) = \sqrt{\frac{2b}{\pi}} e^{-br^2}, 
\end{equation}
with $b$ being the trial parameter and $R=1/\sqrt{b}$ being the polaron size, where we assume $R{\gtrsim}\,a_{B}$.
The polaron energy reads
\begin{eqnarray}
\label{dEtot:1}
&&\delta E_s = \frac{\hbar^2b}{m} - \\
&&\frac{D^2}{8\varrho \pi \varkappa \omega_0} \left\{2{\rm Ci}(\xi)\sin{\xi} +  {[\pi - 2{\rm Si}(\xi)]\cos{\xi}}\right\}, \nonumber
\end{eqnarray}
where $\xi = \omega_0/(4b \varkappa)$, and
\begin{equation}
{\rm Ci}(\xi) = -\int_\xi^\infty \frac{\cos{t}}{t}\, dt, \quad {\rm Si}(\xi) =\int_0^\xi \frac{\sin{t}}{t} \, dt.
\end{equation}
Equation~\eqref{dEtot:1} has a minimum only at sufficiently large $\beta \gg 1$. Performing optimization of the polaron energy we obtain 
\begin{equation}
\label{polaron:shift:strong:kappa}
\delta E_s =  - \frac{D^2}{8\varrho \varkappa \omega_0}.
\end{equation}
This result is parametrically larger than the weak-coupling asymptotics, Equation ~\eqref{pert:2}, valid at $\beta \ll 1$. 
We can also evaluate the polaron mass~\cite{landau:pekar} (see Appendix \ref{app:mass} for another approach) as:
\begin{eqnarray}
\label{mass:strong:1}
&\displaystyle{\frac{m^{*}}{m}}  =  1+ \sum_{\bm q}  \displaystyle{\frac{f_{\bm q}^2}{\varrho \omega_q^4}{\frac{q^2}{m}} {\approx}} \\ 
&\displaystyle{\frac{\beta}{{8}\pi}}
\left(\displaystyle{\frac{\mathcal K}{\varkappa}}\right)^2 %\times \\
%&&
\left\{2 - \xi_{0} \left[ 2{\rm Ci}(\xi_{0})\sin{\xi_{0}} +  {[\pi - 2{\rm Si}(\xi_{0})]\cos{\xi_{0}}}\right]\right\},\nonumber
\end{eqnarray}
where $\xi_{0}$ corresponds to the minimum of Equation ~\eqref{dEtot:1}. Particularly, at $\mathcal K/\varkappa \gg 1$ we 
can put $\xi_{0}\to 0$ with the result 
\begin{equation}
\label{mass:strong:2}
m^{\ast}=\frac{\beta}{{4}\pi}\left(\frac{\mathcal K}{\varkappa}\right)^2 m \gg m.
\end{equation}

In order to study the transition between the limits of small and large coupling 
constants $\beta$ we have performed Feynman variational calculation of the 
polaron energy and mass~\cite{PhysRev.97.660}. To simplify the calculation and 
avoid taking into account full phonon dispersion in the form of Equation ~\eqref{2ML:omega} we assumed 
dispersionless phonons and introduced the smooth cut-off of the exciton-phonon matrix element as
\begin{equation}
\label{smooth:Uq}
|U_q|^2 = \frac{\hbar D^2}{\varrho \omega_0 {\mathcal S}} \exp{\left(-\frac{q^2}{Q^2} \right)}.
\end{equation}
Note that in the strong coupling limit, corresponding to formation of well-localized polaron state, the cut-off $Q$ roughly corresponds 
to the transition to parabolic dispersion of 
phonons, $Q\sim \sqrt{\omega_0 /\varkappa}$ {or to the inverse Bohr radius of exciton, see discussion above.}

%\begin{widetext}
\begin{figure*}[t]
\includegraphics*[width=0.9\textwidth]{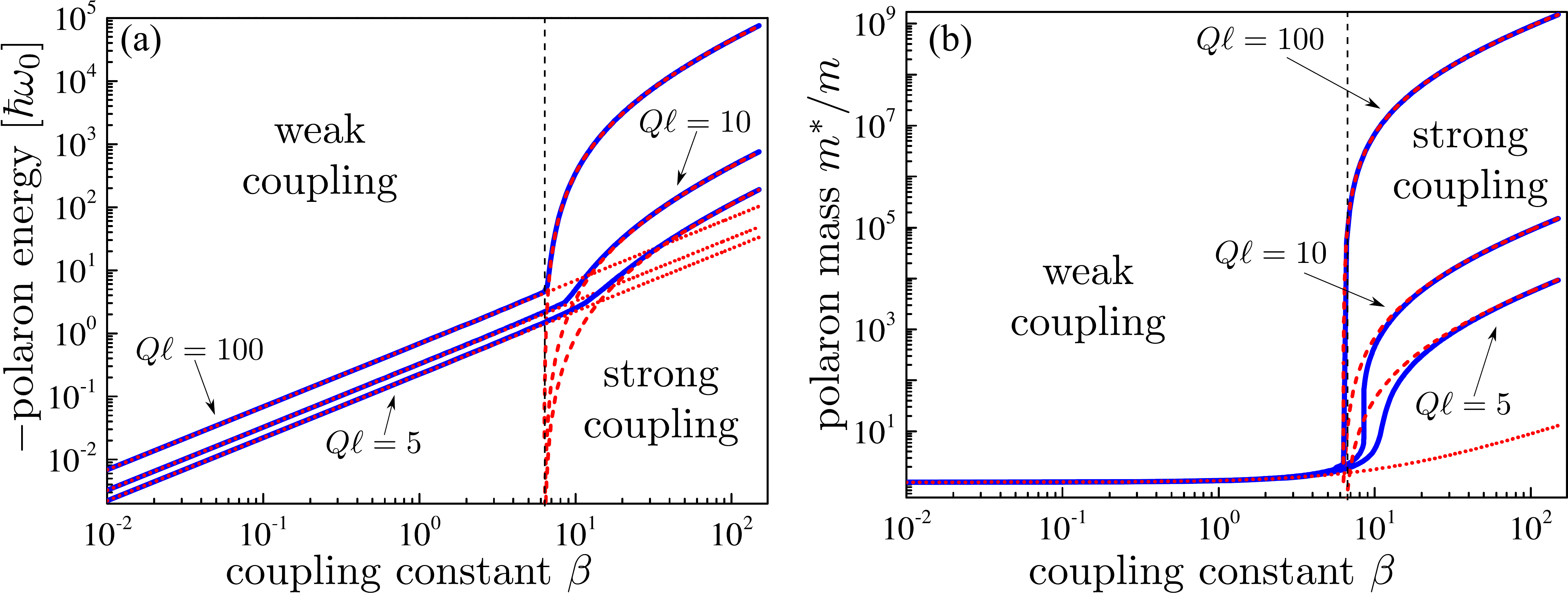}
\caption{Absolute value of the negative polaron energy (a) and the effective mass (b) as functions of the coupling parameter $\beta$ for 
different values of the cut-off parameter $Q\ell,$ marked near the plots. 
Solid blue lines show Feynman variational calculation, dotted and dashed red lines show the analytical 
expressions for Equations  ~\eqref{weak:cutoff} and \eqref{Es:gauss:smooth} valid at, respectively, 
small and large $\beta$. The presented plots, demonstrating a steep increase at $\beta>2\pi,$ are in agreement with the analytical equations.}
\label{fig:pol}
\end{figure*} 
%\end{widetext}

Following \cite{PhysRev.97.660} and supplement to Ref. \onlinecite{PhysRevB.100.041301} we formulate the variational approach 
expressing the kinetic $(E_{\rm kin})$ and potential $E_{\rm pot}$ energies of the polaron with the trial parameters $w$ and $v$ as:
\begin{equation}
\label{K:Feyn:smooth}
{E_{\rm kin}}=\hbar\omega_0 \frac{(v-w)^2}{2v},
\end{equation}
\begin{equation}
\label{P:Feyn:smooth}
{E_{\rm pot}}= -\hbar\omega_0 \frac{\beta}{4\pi} \int_0^\infty \frac{e^{-u} {du}}{(Q\ell)^{-2} +2 F},
\end{equation}
with $\ell = \sqrt{\hbar/(2m\omega_0)}$ being the oscillator length corresponding to the bare exciton mass, 
and auxiliary function
\begin{equation}
F=\frac{w^2}{2v^2}u+\frac{v^2-w^2}{2v^3}\left(1-e^{-uv}\right).
\label{Ffunction}
\end{equation}
{Figure \ref{fig:pol}} shows the polaron energy calculated using Feynman's variational 
method by optimization of ${E_{\rm kin}+E_{\rm pot}}$ vs. trial parameters $u$ and $v$ (solid blue line) and the 
asymptotic dependence in the weak and strong coupling regimes (dotted and dashed red lines, respectively). 
The weak coupling energy shift depends on the cut-off parameter logarithmically while the mass is independent of the cut-off
\begin{equation}
\label{weak:cutoff}
\delta E_w = -\hbar\omega_0 \frac{\beta}{2\pi} \ln{(Q\ell)}, \quad m^*=m\left(1+\frac{\beta}{4\pi}\right).
\end{equation}
This result exactly matches the calculation, Equations \eqref{pert:2} and \eqref{pert:M}, with dispersive 
phonons at $Q=\sqrt{\omega_0/\varkappa}$. At the strong coupling the model with the cut-off produces the polaron shift and mass at $\beta> 2\pi$:
\begin{eqnarray}
\label{Es:gauss:smooth}
\frac{\delta E_s}{\hbar\omega_0} = 
- \frac{(Q\ell)^2}{4\pi}\left(\sqrt{\beta}-\sqrt{2\pi} \right)^2, 
\\ \frac{m^*}{m} = \frac{(Q\ell)^4}{2\pi}(\sqrt{\beta}-\sqrt{2\pi})^2. \nonumber
\end{eqnarray}
At $Q=\sqrt{\omega_0/\varkappa}$ the {energy} in Equation \eqref{Es:gauss:smooth} differs by a 
factor $2/\pi$ from the calculation with the full dispersion given by Equation ~\eqref{polaron:shift:strong:kappa}, 
while the effective mass differs by a factor of $2$ from Equation \eqref{mass:strong:1}. 
The Feynman approach demonstrates that the polaron energy almost smoothly varies 
between the weak and strong coupling. Thus, it is anticipated that the energy of the polaron 
in the model with the full dispersion will also smoothly pass from the weak to the strong 
coupling asymptotes, Equations  ~\eqref{pert:2} and \eqref{polaron:shift:strong:kappa} with the 
transition at $\beta \sim {2\pi}$. Inset in Figure ~\ref{fig:pol} demonstrates the polaron mass $m^{\ast}/m$ 
as a function of the dimensionless coupling constant $\beta$ calculated in Feynman's approach (solid blue line) 
and its asymptotes. Interestingly, the effective mass abruptly increases at $\beta \sim 2\pi$ 
where the transition between the weak and strong coupling regimes is achieved corresponding to a fast compression
of the polaron state with the increase of $\beta$ in this domain and this qualitative feature does not depend on the trial function.
Similarly to the energy, the polaron mass quite accurately follows the asymptotes 
obtained in the weak and strong coupling regimes.

\section{Discussion: experimental consequences and related effects}

Now we can discuss possible relation of our theoretical results to the experimentally 
investigated bilayer materials and the effects of exciton-polarons on their properties in different coupling regimes.
We begin with the estimates of the coupling constant $\beta$ in \eqref{beta} and corresponding energy shift in 
\eqref{pert:2}. Since the variety of the structures and structure-related  
properties of the excitons is very broad, for the theoretical analysis 
we accept the typical system parameters $\varepsilon\approx 5$ and $a_{B}\approx 1\mbox{ nm}$ and 
estimate the deformation potential as $D\approx 1.5\times 10^{3}\mbox{ meV/nm}.$ {To estimate $\beta,$ 
we take the van der Waals-related phonon energy $\hbar\omega_{0}=3\mbox{ meV},$ when the MLs are 
in contact with each other ~\cite{Jeong2016,PhysRevB.91.165403,Goldstein:2016aa}, corresponding to the 
estimate in subsection \ref{sub:weak}  for $L\sim 2a_{0}$.}
Taking into account that $M\sim 5\times\,10^{5}m_{0},$
$2m\sim\,m_{0},$ and the unit cell area $a_{0}^{2}$ is close to $0.1\mbox{ nm}^{2},$
we obtain for the typical structures where the distance between the layers is of the order 
of $a_{0},$ the values of $\beta$ of the order of $5\times 10^{-2}$.
For the ratio ${\cal K}/{\varkappa}\sim\,10^{3},$ the corresponding shift $\delta E_{w}$ 
at temperature $T=100\mbox{ K}$ is then of the order of $0.3\mbox{ meV}.$
At a constant $D$ this value of $\beta$ can be enhanced by increasing the interlayer distance since van der Waals vibration 
frequency behaves as $\omega_{0}^{2}\propto 1/L^{6}$, and the 
shift $\delta E_{w}$ behaves at $T=0$ as $\propto L^{3}$ then. The role of the interlayer distance dependence on the deformation potential constant $D$ 
can be characterized as follows. At $L\leq a_{B},$ $D$ is weakly $L-$dependent, and, therefore $\beta\propto L^{6}$. At larger 
distances, where $D\propto L^{-2}$ one obtains $\beta\propto L^{2}.$ The temperature enhances the value of $\beta$ by the 
factor $\propto T L^{3}.$ The bilayers with large values of $L$ can be designed similarly to the drum-like structures 
of Refs. \onlinecite{Morell:2016aa,Morell:2019aa}. 

{It is noteworthy that the enhancement of the coupling constant with increasing interlayer distance is 
mainly due to the reduction of the interlayer interaction characterized by the diminishing $\omega_{0}$. 
The smaller $\omega_{0}$ is, the softer is the phonon dispersion and the larger is $\beta$. Formally, in the 
limit of isolated layers $\beta\to \infty$. However, the physical quantity $E_{w}$ in Equation  \eqref{pert:1:T} at 
$T=0$ decreases as $\beta\omega_{0}\propto 1/L.$ In addition, the finite size of the TMDC flakes results in 
quantization of phonon wavevectors, which cuts off dispersion ~\eqref{1ML:omega} at small $q\sim 1/L_{s}$ 
where $L_s$ is the flake size. Thus, in real samples, the quantity $\omega_{s}\sim\varkappa/L_{s}^2$ represents a
natural cut-off frequency and should be used in the denominator of $\beta$ if $L\gtrsim L_{s}$. As a result, 
for $L\gtrsim L_{s},$ the polaron coupling constant starts to decrease with increasing $L$ due to reduction of the deformation potential constant $D$.}

{In the considerations above we have disregarded the moir\'e effects in bilayer structures, which can be 
important when the layers are in a relatively close contact. The resulting appearance 
of the superstructure potential due to the twist of the lattices and their incommensurability on the properties 
of the interlayer polarons, should be analyzed self-consistently for both phonons and excitons.  
This is a problem for the future study.}

In the strong coupling regime the polaron effects produce a Stokes shift energy between the absorption and emission.
{The polaron effects can also influence the transport of excitons, their diffusion, the phonon-related drag, and Raman scattering since 
the exciton is surrounded by a phonon cloud. These effects are now actively studied in 
TMDC-based and other layered nanosystems~\cite{PhysRevLett.120.207401,PhysRevB.100.045426,2019arXiv191002869Y,2020arXiv200201561L,Reichardt2020}.}

Finally, we discuss collective properties of these polaronic systems. First, lattice deformation and phonon 
clouds surrounding the exciton, will influence the {interactions between two excitons.} For example, at large interlayer 
distances due to strong coupling the displacements can be sufficiently large that the layers may start touching each other
leading to formation of the stronger bound excitons.  At the concentration of carriers per layer $n_{2D},$ 
these single-exciton states can be formed at $BL^{2}n_{2D}\le E_{B},$ where $E_{B}$ is the exciton binding energy,
corresponding to the balance of the elastic energy loss {per single electron-hole pair} ($BL^{2}n_{2D}$) and the energy gain ($E_{B}$)
due to formation of the exciton. {Note that this discussion is valid provided that $n_{2D}$ is below the 
Mott transition density~\cite{RevModPhys.90.021001} which can be roughly estimated as $n_{\rm M} \sim 1/a_{B}^{2}$ at $a_{B}\gg L$ 
and $n_{\rm M} \sim 1/a_{B}^{2}\times(a_{B}/L)^{3/2}$ at $a_{B}\ll L$ (see Appendix \ref{app:form-factor}).} 
At higher $n_{2D},$ these polarons {overlap} and can form macroscopic
many-body states similar to the fluctuons near the phase transitions \cite{Krivoglaz1974}, where the host lattice is soft
due to the vicinity of the transition, while in the bilayers of our interest the softness is due to 
the flexural character of the elasticity. The polaron effect, enhancing exciton-exciton interactions 
can be also important for realization of condensed excitonic phases or 
electron-hole condensates in strong magnetic fields.
 
A similar polaron effect can also strongly affect impurity states in suspended monolayer semiconductors. 
For example, if a positively charged impurity is placed in the vicinity of a monolayer,  
the electron attraction to the impurity will be accompanied by the monolayer deformation affecting the bound states energies 
and wavefunctions. The effect can be most pronounced and enhanced by the corresponding nonuniform polarization of the
layers for close-to-critical or supercritical Coulomb impurities in narrow-gap systems, 
including gapless graphene~\cite{shytov:236801,shytov:246802}. 
The predicted polaron effects can also be important for 
quantitative description of the bound states in Dirac layered materials~\cite{Downing:2017aa}.

\section{Conclusion}

We studied theoretically possible formation of interlayer exciton-polarons in two-layer structures 
of transition metal dichalcogenides. The polaron is formed 
due to the Coulomb attraction between electron and hole located in different layers, 
which results in the out-of-plane deformations of the individual layers. The latter are 
related to the flexural phonons which are extremely soft in two-dimensional materials.
This electron-hole attraction decreases the 
interlayer distance with the effect dependent on the electric field inside the exciton, coupling between the 
layers due to the van der Waals forces, and the flexural rigidity of the layers. This, a polaron is being formed.
Since the van der Waals-determined frequencies of the coupled layers vibrations are usually low, the temperature can have a large effect
on the polaron-related effects. The total energy 
and effective mass of such exciton-polaron depend of these parameters. In the strong coupling limit we applied 
two forms of the variational approaches, including the Fr\"{o}lich and Feynman variational procedures.
The effective mass rapidly increases when the exciton-phonon coupling exceeds a critical value, 
corresponding to a spatial compression of the polaron state. Taking into account that the characteristics
of excitons in the transition metal dichalcogenides are strongly preparation-dependent,
we expect different realizations of the polaron-related effects in their experimentally observable properties.

\section*{Acknowledgements}
M.A.S. acknowledges partial support of the Russian Science Foundation project \# 19-12-00273 (numerical variational calculations). 
M.M.G. is grateful to the Russian Science Foundation project \# 19-12-00051 for partial support (analytical theory).
E.S. acknowledges support of the Spanish Ministry of Science and the European Regional Development
Fund through PGC2018-101355-B-I00 (MCIU/AEI/FEDER, UE) and the Basque Country Government through
Grant No. IT986-16.

\begin{appendix}

\section{Matrix elements and form-factor of exciton-phonon coupling}
\label{app:form-factor}
 
Here we derive the matrix elements of interaction of the interlayer exciton with the out-of-plane vibrations. 
To that end we introduce the electron-hole interaction potential, $V(\bm \rho_{e}-\bm \rho_{h},z_{e} - z_{h})$, 
as a function of the in-plane coordinates, $\bm \rho_{e,h}$, and out-of plane coordinates, $z_{e,h}$. 
In the absence of phonons $z_{e} - z_{h} = L$, the interlayer distance.

The exciton wavefunction takes the form of Equation ~\eqref{exc:wave}, with
\begin{equation}
\bm R = \frac{m_{e}}{m} \bm \rho_{e} + \frac{m_{h}}{m} \bm \rho_{h},\qquad
\bm \rho = \bm \rho_{e} - \bm \rho_{h},
\end{equation}
being the in-plane center of mass and relative coordinates, respectively.

The matrix element of the exciton interaction with the out-of-plane phonons results from the 
variation of the exciton energy in the lowest order in the layer displacements 
$\zeta_{1}$ and $\zeta_{2}$ can be written with $V_{z}(\rho,z)\equiv{\partial V(\rho,z)}/{\partial z}$ as:
\begin{equation}
\label{U:rspace}
U_{\bm K'\bm K}=\hspace{-0.2cm}\int\,\hspace{-0.1cm}d^{2}Rd^{2}\rho\Psi_{\bm K'}^{*}\Psi_{\bm K} 
V_{z}(\rho,{L})[\zeta_{1}(\bm \rho_{e}) - \zeta_{2}(\bm \rho_{h})],
\end{equation}
where the electron and the hole reside in the top (1) and the bottom (2) layer, respectively,
as shown in Figure  \ref{fig:scheme}. Here we are interested only in the interaction 
with the even-symmetry mode, which provides the maximum coupling, while 
the contribution of the odd-symmetry vibrations vanishes after averaging with 
axially symmetric envelope of relative motion.

Performing quantization of displacements after Equation ~\eqref{displ:qnt} and 
integrating in Equation ~\eqref{U:rspace:1} over $\bm R$ we arrive at
\begin{eqnarray}
\label{U:kspace}
U_{\bm q} = \frac{1}{2}
{\sqrt{\frac{\hbar}{\varrho \omega_q \mathcal S}}}
 \int d^{2}\rho\,|\varphi(\rho)|^2 {V_z(\rho,L)}  \\
\times\left[\exp{\left(-\mathrm i \bm q \bm \rho \frac{m_{h}}{m}\right)} + \exp{\left(\mathrm i \bm q \bm \rho \frac{m_{e}}{m}\right)}\right] 
+ {\rm h.c.}
\nonumber
\end{eqnarray}

We describe electron-hole interaction by the Coulomb law, i.e., we neglect an intricate screening in a bilayer system [cf. Ref. \onlinecite{Semina:2019aa}]:
\begin{equation}
\label{Vc}
V(\rho,z) = -\frac{e^2}{\varepsilon} \frac{1}{\sqrt{\rho^2+z^2}},
\end{equation}
with the $z-$ derivative
\[
{V_z(\rho,L)=}\left.\frac{\partial V(\rho,z)}{\partial z}\right|_{z=L} 
%=\left.\frac{\partial V(\rho,\zeta)}{\partial \zeta}\right|_{\zeta=0}
= \frac{e^2}{\varepsilon}\frac{L}{(\rho^2+L^2)^{3/2}}. 
\]
We begin with realization $L\ll a_{B},$ where we may take the hydrogen-like wavefunction of the exciton 
\begin{equation}
\varphi\left(\bm{\rho}\right) =\sqrt{\frac{2}{\pi a_B^2}}e^{-\rho/a_{B}},
\end{equation}
and it is convenient to present the matrix element of the exciton-phonon coupling as
\begin{equation}
\label{U:kspace:small}
U_q = {\sqrt{\frac{\hbar}{\varrho \omega_q \mathcal S}} \frac{4e^2}{\varepsilon a_{B}^{2}} \mathcal F_s(q)}. 
\end{equation}
Performing integration in Equation  \eqref{U:kspace} and taking into account that in the limit 
of $L \ll a_{B}$ the exponent is inessential  
since the integrand has singularity at $\rho\to 0$ at $L=0,$ and the integration can be performed analytically with the result, and 
taking into account the definition in \eqref{classical:int}, we obtain
\begin{equation}
\label{FF:small:1}
 \mathcal F_s(q) = \frac{1}{2} \left(e^{-q L m_{e}/m} + e^{-q L m_{h}/m}\right).
\end{equation}
Since $L$ in this realization is of the order of few units cell sizes $a_{0}$, the corresponding wavevector $1/L \gg \sqrt{\omega_{0}/\varkappa},$ that is the 
size integration domain in Equation  \eqref{pert:1:T}, we can safely put $F_{s}(q)=1$ there.

Note that the behavior of exciton-phonon coupling at $q=0$ can be understood by considering the 
exciton as a thin nanosize capacitor with nonuniform charge density, where we introduce the probability 
densities
\begin{equation}
n_{h}(r)=\frac{4}{a_{h}^{2}}\frac{e^{-2r/a_{h}}}{2\pi};\qquad n_{e}(r)=%
\frac{4}{a_{e}^{2}}\frac{e^{-2r/a_{e}}}{2\pi }, 
\end{equation}
where $a_{h} =a_{B}{m_{e}}/{m}$ and $a_{e}=a_{B}-a_{h}=a_{B}m_{h}/{m}$
for hole and electron, respectively. These distributions produce electric fields
\begin{eqnarray}
\mathcal{E}_{h}(r)=2\pi\frac{e}{\varepsilon} n_{h}(r)=\frac{4}{a_{h}^{2}}e^{-2r/a_{h}} \\
\mathcal{E}_{e}(r)=-2\pi\frac{e}{\varepsilon} n_{e}(r)=-\frac{4}{a_{e}^{2}}e^{-2r/a_{e}}. \nonumber
\end{eqnarray}
The function $f({\bm r})$ in Equation  \eqref{1ML:-} becomes $\left|f({\bm r})\right|=e\left({\mathcal{E}}_{h}(r)n_{e}(r)-{\mathcal{E}}_{e}(r)n_{h}(r)\right)$
with the corresponding Coulomb force between the electron and the hole:
\begin{eqnarray}
\label{eq:F}
&&|{\mathcal F}_{C}|=\frac{1}{2} \int \left|f({\bm r})\right| d^{2}r \\
&&=e\int\mathcal{E}_{h}(r)n_{e}(r)d^{2}r =-e\int\mathcal{E}_{e}(r)n_{h}(r)d^{2}r. \nonumber
\end{eqnarray}
After integration in Equation  \eqref{eq:F} we obtain
\begin{equation}
|{\mathcal F}_{C}|=\frac{e^{2}}{\varepsilon}\frac{4}{a_{B}^{2}},
\end{equation}
in agreement with the full calculations.

For completeness, we consider the opposite limit $L \gg a_{B},$ where the 
{the potential of electron-hole interaction can be approximated by harmonic potential: 
$V(L,\rho)\approx -e^{2}(1+\rho^{2}/2L^2)/(\varepsilon L).$ The resulting}  
relative motion of electron and hole is described 
by a Gaussian-like function:
\begin{equation}
\varphi \left( \bm{\rho}\right) =\sqrt{\frac{1}{\pi \tilde{a}_{B}^{2}}}e^{-\rho^{2}/2\tilde{a}_{B}^{2}},
\end{equation}
where $\tilde{a}_{B}={a}_{B}\times\left(L/{a}_{B}\right)^{3/4}\gg a_{B}.$  Thus, we can disregard $\rho^{2}$ 
in the denominator of Equation ~\eqref{Vc} and obtain, using definition of $D$ in Equation  \eqref{classical:int}: 
\begin{equation}
 \mathcal F_s(q) = \frac{1}{2}
 \left(
 e^{-(\tilde{a}_{B}qm_{e}/(2m))^{2}} + e^{-(\tilde{a}_{B}qm_{h}/(2m))^{2}}
 \right).  
\label{FF:large} 
\end{equation}

Now we analyze the role of the mass-related factors in the exponents. While typically in TMDC-based systems the electron 
and hole masses have similar values (see Refs. \onlinecite{PhysRevB.100.041301,Goryca:2019aa}), it is instructive 
to address the case where $m_{e}$ and $m_{h}$ 
are considerably different. Let us assume that $m_{h}\ll m_{e},$
then $m_{e}\approx m$ and the phonon causes displacement of the heavier
particle, that is, the electron corresponding to the first term
in the square brackets in \eqref{U:rspace}. Then almost the entire phonon 
momentum [the fraction ${\sim}\left(m_{{e}}/m\right) $ of
it] will be absorbed in the center-of-mass motion, and, therefore, will not
contribute into variation of the electron-hole Coulomb interaction. Only the residual small
part [${\sim}\left(m_{{h}}/m\right)$ of the phonon momentum] 
will affect the relative position of 
electron and hole and, thus, produce the form-factor. For this reason, a
much larger momentum is required to produce a considerable change in the Coulomb energy. Similar analysis 
can be applied for the contribution of the lighter particle, that is the hole.

\section{Effective mass of the polaron: slow motion of the monolayers}
\label{app:mass}

Here we consider a double-layer exciton-polaron characterized by $z-$axis
displacement in each monolayer 1 (2) as $\zeta_{1}\left({\bm r}\right)=-\zeta_{2}\left({\bm r}\right),$
where ${\bm r}$ is the in-plane coordinate and $\zeta\left({\bm r}\right)=2\zeta_{1}\left({\bm r}\right)$. 
Let it move slowly with the velocity ${\bm v}$ such that the time-dependent displacement 
becomes $\zeta_{1}\left( {\bm{r-v}}t\right).$ Kinetic energy of the polaron (moving deformation) is then 
\begin{equation}
{E_{\rm K}} =\frac{\varrho }{2}\times\frac{1}{2}\int \left( \frac{\partial \zeta \left( 
{\bm{r-v}}t\right) }{\partial t}\right) ^{2}d^{2}r
\equiv m^{\ast }\frac{v^{2}}{2}, 
\end{equation}
where $m^{\ast}$ is the polaron effective mass with $m^{\ast}\gg m$ in this strong-coupling regime.
The low-$v$ expansion
\begin{equation}
\frac{\partial \zeta \left( {\bm{r-v}}t\right) }{\partial t} =-{\bm v}\mathbf{\nabla }\zeta \left( {\bm r}\right),
\end{equation}
yields
\begin{equation}
{E_{\rm K}} =\frac{1}{4}{\varrho }\int \left({\bm v}\mathbf{\nabla }%
\zeta \left( {\bm r}\right) \right) ^{2}d^{2}r.
\end{equation}
Introducing unit vector $\bm{n}\equiv{\bm v}/v,$ we obtain
\begin{equation}
{E_{\rm K}} =\frac{1}{4}{\varrho v^{2}}\int \left( \bm{n}\mathbf{\nabla }\zeta \left( {\bm r}\right) \right) ^{2}d^{2}r,
\end{equation}
resulting in the polaron mass 
\begin{equation}
\label{mast}
m^{\ast} =
\frac{\varrho}{2}\int \left( \bm{n}\mathbf{\nabla}\zeta \left({\bm r}\right) \right) ^{2}d^{2}r=
\frac{\varrho}{4}\int \left(\mathbf{\nabla}\zeta \left( 
{\bm r}\right) \right) ^{2}d^{2}r.
\end{equation}
This definition fully corresponds to Equation ~\eqref{mass:strong:1} since
\begin{equation}
\sum_{\mathbf{q}}\frac{q^{2}f_{\mathbf{q}}^{2}}{\varrho \omega _{q}^{4}}=
\frac{\varrho}{4}\sum_{\mathbf{q}}q^{2}\zeta _{\mathbf{q}}^{2},
\end{equation}%
is equation \eqref{mast} in the momentum representation. 
Now we can estimate the effective mass taking into account that 
for a monotonic polaron function $\zeta\left(r\right)$ 
\begin{equation}
\int \left( \bm{n}\mathbf{\nabla }\zeta({\bm r})\right)^{2}d^{2}r \sim \zeta ^{2}(0). 
\end{equation}
Thus, we arrive at the relation between $m^{\ast}$ and the maximum displacement $\zeta(0):$
\begin{equation}
m^{\ast }\sim \varrho \zeta^{2}(0).
\end{equation}
Taking into account the condition of linear exciton-phonon coupling
$\zeta(0)\ll L$, we obtain the restriction $m^{\ast }\ll \varrho L^{2}.$ 

\end{appendix}

%\bibliographystyle{MSP}
%\bibliography{all1}

\end{document}